\documentclass{webofc}

\usepackage[varg]{txfonts}   
\usepackage{hyperref}
\usepackage{url}
\hypersetup{colorlinks=true,citecolor=blue,urlcolor=blue,linkcolor=blue}
\begin{document}
\title{Quarkonia Theory: From Open Quantum System to Classical Transport}
%
%

\author{\firstname{Xiaojun} \lastname{Yao}\inst{1}\fnsep\thanks{\email{xjyao@uw.edu}} 
}

\institute{InQubator for Quantum Simulation, Department of Physics, University of Washington, Seattle, WA 98195, USA.
}

\abstract{This is a theoretical overview of quarkonium production in relativistic heavy ion collisions given for the Hard Probes 2024 conference in Nagasaki. The talk focuses on the application of the open quantum system framework and the formulation of the chromoelectric correlator that uniquely encodes properties of the quark-gluon plasma relevant for quarkonium dynamics and thus can be extracted from theory-experiment comparison.
}
\maketitle
\section{Introduction}
\label{intro}
Good afternoon. I was asked to review quarkonium theory. Our conference is called Hard Probes. I would like to start with one question: what is the unique property of the quark-gluon plasma (QGP) that we are probing via quarkonium?

Quarkonium has been used to probe deconfinement in heavy ion collisions since the early studies of plasma screening effect and quarkonium melting in a hot medium~\cite{Matsui:1986dk,Karsch:1987pv}. 
In this conference, we have seen many new exciting experimental data on quarkonium nuclear modification factor $R_{AA}$ and the elliptic flow coefficient $v_2$, which are affected by many physical factors such as hot and cold medium effects and feed-down contributions in the hadronic stage. We want to use the hot medium effect on the $R_{AA}$ and $v_2$ to probe properties of the QGP, which is not a simple task. As I will show you, it involves theory developments, phenomenological calculations and computational advancements.

Many in the audience may have their own answers to the question I raised. Let me be more specific here: We will focus on the experimental data at low transverse momentum, e.g., $p_T$ below the quarkonium mass. There are at least two benefits of focusing on the low-$p_T$ region. First, it allows us to treat the heavy quark-antiquark ($Q\bar{Q}$) pairs in a nonrelativistic way. Second, it means that the hot medium only affects the evolution from unbound $Q\bar{Q}$ pairs to final bound quarkonium states, but not the initial production of the $Q\bar{Q}$ pairs, which can be well calculated via perturbative QCD methods with nuclear parton distribution functions (nPDF) accounting for some of the cold medium effects. At high $p_T$, one also needs to consider the medium effects on the production of the $Q\bar{Q}$ pair from the gluon splitting $g\rightarrow Q\bar{Q}$. In this conference, there is already a talk on the high-$p_T$ case that focuses on $J/\psi$ production within a jet~\cite{zhang}.

In this talk, I will show you how the low-$p_T$ experimental data can probe a physical object called the chromoelectric correlator~\cite{Yao:2020eqy}. It sounds abstract. Physically, it gives novel generalized transport coefficients and new types of generalized gluon distributions in the medium that are relevant for quarkonium dynamics. Going from left (the experimental data) to right (the chromoelectric correlator) is not an easy task. We need some theoretical tools to guide us. They are the open quantum system framework and effective field theory.

\section{Theoretical Tools}
\label{sec-1}
\subsection{Open Quantum System}
\label{sec-1-1}
The open quantum system framework is widely used in quantum optics and quantum information science. It was first introduced in the heavy ion field to study quarkonium by Akamatsu and Rothkopf~\cite{Akamatsu:2011se,Akamatsu:2014qsa}. The open quantum system framework treats $Q\bar{Q}$ pairs and the QGP together as a closed quantum system $\rho_{\rm tot}$, evolving unitarily. By just focusing on the degrees of freedom of the $Q\bar{Q}$, we end up with a non-unitary and time-irreversible evolution equation
\begin{align}
\rho_{Q\bar{Q}}(t) = {\rm Tr}_{\rm QGP}[ U(t) \rho_{\rm tot}(0) U^\dagger(t) ] \,.
\end{align}
In certain limits, the evolution equation can be written as a Lindblad equation, which is a non-unitary generalization of the Schr\"odinger equation and is Markovian, trace-preserving and completely positive. Its semiclassical limit gives various Boltzmann equations. Since the early work, there have been a lot of open quantum system studies for quarkonium, many of which combine this framework with nonrelativistic effective field theories~\cite{Brambilla:2016wgg,Blaizot:2017ypk,Yao:2018nmy,Miura:2019ssi}. Here I listed four reviews~\cite{Rothkopf:2019ipj,Akamatsu:2020ypb,Sharma:2021vvu,Yao:2021lus}, each covering a different aspect. 

\subsection{Effective Field Theory}
\label{sec-1-2}
\begin{figure*}
\centering
\vspace*{0cm}       
\includegraphics[width=12cm,clip]{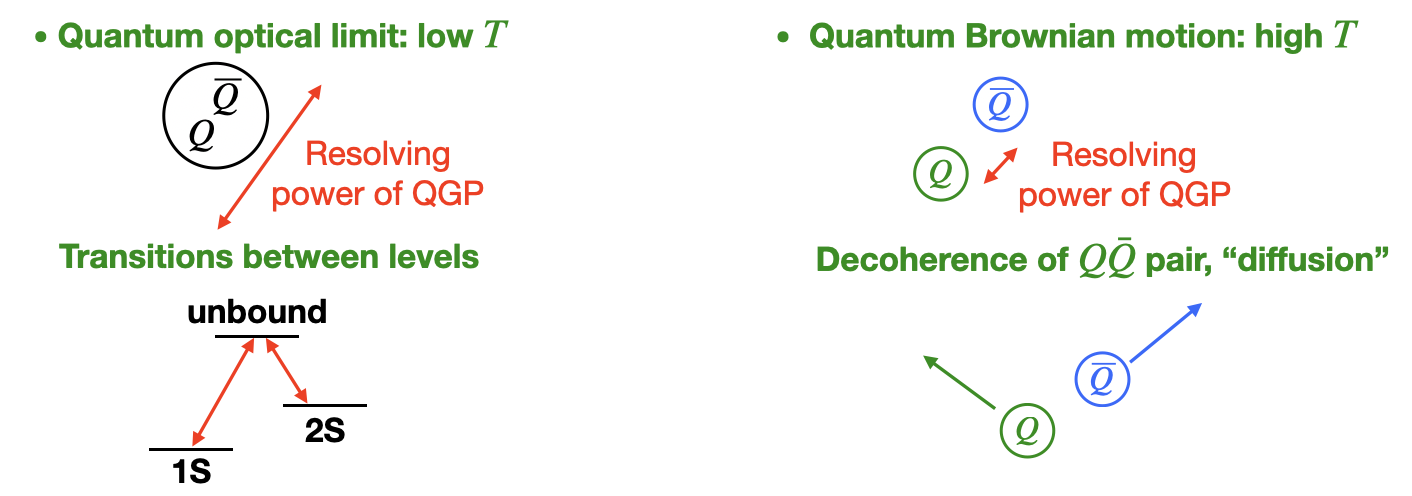}
\caption{Quantum optical (left) and quantum Brownian motion (right) limits for the effective descriptions of quarkonium dynamics in the medium.}
\label{fig:eft}
\end{figure*}

The second tool is effective field theory (EFT), which is nothing but separation of scales. For quarkonium states in vacuum, we have three energy scales: the heavy quark mass $M$, the inverse of the quarkonium size $\frac{1}{r}$ and the binding energy $E_b>0$. For charm and bottom systems, these scales are listed here in GeV: $M\sim1.5$, $\frac{1}{r}\sim0.9$, $E_b\sim0.5$ for charm and $M\sim4.5$, $\frac{1}{r}\sim1.5$, $E_b\sim0.5$ for bottom. By integrating out high energy scales, we can obtain different EFTs such as nonrelativistic QCD (NRQCD)~\cite{Bodwin:1994jh} and potential NRQCD (pNRQCD)~\cite{Brambilla:1999xf}. This way of constructing an EFT is called the top-down approach. In thermal medium, we have thermal scales such as the medium temperature $T$ and the Debye mass $m_D$. For simplicity, I will use $T$ to represent them all. Depending on where $T$ fits, we can have different descriptions as shown in Fig.~\ref{fig:eft}. When $T$ is small, the resolution power of the QGP is weak and the QGP interacts with the $Q\bar{Q}$ as a whole. In this case, the effective dynamics is best described by transitions between different levels: bound and unbound. This limit is called the quantum optical limit due to the similarity with quantum optics. At high temperature, the resolution power is strong and the QGP interacts with each individual heavy quark independently. In this limit, the heavy quark and antiquark are jiggling in space, leading to decoherence of their wavefunction. Because of this similarity with diffusion, this limit is called the quantum Brownian motion limit.

The separation of energy scales can be summarized in Table~\ref{tab:scale}. In the first case, $M\gg T \gg E_b,\Lambda_{\rm QCD}$, the EFT that is useful is the NRQCD. The quantum description of the time evolution is a Lindblad equation in the quantum Brownian motion limit. Its classical limit corresponds to a diffusion equation~\cite{Blaizot:2017ypk}. In the second case, $M\gg\frac{1}{r}\gg T,\Lambda_{\rm QCD}$; $T \gg E_b$, the EFT useful is the pNRQCD. The quantum description of the time evolution is also a Lindblad equation in the quantum Brownian motion limit~\cite{Brambilla:2016wgg}. Its classical limit has not been worked out but should be similar to a diffusion equation. In the third case, the scale hierarchy is $M\gg \frac{1}{r} \gg T,E_b,\Lambda_{\rm QCD}$ and the relevant EFT is the pNRQCD. Whether a Lindblad equation exists in this case is unknown, but a classical description is known in the quantum optical limit, which is a Boltzmann equation~\cite{Yao:2018nmy,Yao:2020eqy}. In the first two cases, one expands $\frac{E_b}{T}$ so only the zero frequency limit of the relevant QGP correlator contributes. In the last two cases, one expands $rT$ but not $\alpha_s$. Therefore some nonperturbative effects can be incorporated. 

\begin{table}
\centering
\caption{Hierarchies of energy scales, EFTs and quantum and classical effective descriptions.}
\label{tab:scale}       
\begin{tabular}{|c|c|c|c|}
\hline
Hierarchy of energy scales & EFT & Quantum description & Classical limit \\\hline
$M\gg T \gg E_b,\Lambda_{\rm QCD}$ & NRQCD & Lindblad equation & Diffusion equation \\\hline
$M\gg r^{-1} \gg T,\Lambda_{\rm QCD}$; $T\gg E_b$ & pNRQCD & Lindblad equation & Unknown \\\hline
$M\gg r^{-1} \gg T,E_b,\Lambda_{\rm QCD}$ & pNRQCD & Unknown & Boltzmann equation \\\hline
\end{tabular}
\vspace*{0cm}  
\end{table}

\section{Various Phenomenological Studies}
In the following, I am going to review different phenomenological models. They use different treatments for each of the three temperature regions I just discussed. For the nomenclature, I will follow Ref.~\cite{Andronic:2024oxz}, which summaries the community's efforts in response to the EMMI Rapid Reaction Task Force. Some people in the audience put great efforts into it. If you have not read it, I highly recommend it.

\subsection{Statistical Hadronization Model}
The first model is the Statistical Hadronization Model~\cite{Andronic:2017pug,Andronic:2019wva}. It assumes heavy quarks reach kinetic equilibrium within the QGP and thus has no non-trivial dynamics in the three temperature regions. The equilibrium is kinetic but not chemical, which is a consequence of the large heavy quark mass. On the freezeout hypersurface, it uses a statistical instantaneous (re)combination model. The model describes $J/\psi$ $R_{AA}$ data well at low $p_T$. At intermediate $p_T$, it underestimates the data, even though the corona effect has been accounted for. The reason is that the thermal $p_T$ spectrum used in the (re)combination model is too soft.

\subsection{Including $\frac{1}{r}$ Scale in Hadronization}
One can introduce the effect of the $\frac{1}{r}$ scale in hadronization by using a Wigner function which knows about the wavefunction of quarkonium. This is the case in the Instantaneous Coalescence Model~\cite{Greco:2003vf} and the Parton-Hadron-String Dynamics model~\cite{Song:2017phm,Song:2023zma}. Coalescence based on the Wigner function has also been applied to quarkonium production in proton-proton collisions, with very intriguing results presented by Gossiaux in this conference~\cite{PB}. In the Wigner function coalescence model the effect of the $E_b$ scale is not included, which has to be done in a dynamical way, i.e., in reaction rates. 

\subsection{Lindblad Equation from NRQCD}
The third example is from the Osaka~\cite{Miura:2022arv} and Nantes~\cite{Delorme:2024rdo} approaches. Both of them focused on this hierarchy of energy scales $M\gg T \gg E_b,\Lambda_{\rm QCD}$ and studied a Lindblad equation in the quantum Brownian motion limit
\begin{align}
\frac{{\rm d} \rho_{Q\bar{Q}}(t)}{{\rm d}t} = -i[H, \rho_{Q\bar{Q}}(t)] + \sum_{ij} D_{ij} \Big(  L_i \rho_{Q\bar{Q}}(t) L_j^\dagger - \frac{1}{2}\big\{ L_j^\dagger L_i, \rho_{Q\bar{Q}}(t) \big\}  \Big) \,.
\end{align}
The Hamiltonian $H$ contains a kinetic term and a temperature-dependent real potential. The Lindblad operators $L_i$ account for transitions between $Q\bar{Q}$ pairs in the color singlet and color octet. Both groups solved the Lindblad equation in one spatial dimension. They differ in the choice of the basis. The Osaka approach used momentum basis so the environment correlator is diagonal $D_{ij}\propto \delta_{ij}$. The Nantes approach used position basis so $D_{ij}$ has off-diagonal parts. They studied static states of the Lindblad equation, i.e., solutions at very late time. The Osaka model found that the occupation probabilities of different Hamiltonian eigenstates follow the Boltzmann distribution for different initial conditions and medium temperatures. The Nantes model studied the evolution of an initially localized singlet state. As time goes on the diagonal part of the density matrix becomes constant (nonzero) at large $r$ in both the color singlet and octet channels, meaning that the system has reached equilibrium of free particles. But at small $r$, the distribution is non-trivial. The singlet channel has a peak while the octet channel has a valley. It is a consequence of the attractive potential in the color singlet channel and the repulsive potential in the color octet channel. The peak corresponds to bound quarkonium states.

\subsection{Boltzmann Equation and pNRQCD}
\begin{figure}[t]
\centering
\includegraphics[width=12cm,clip]{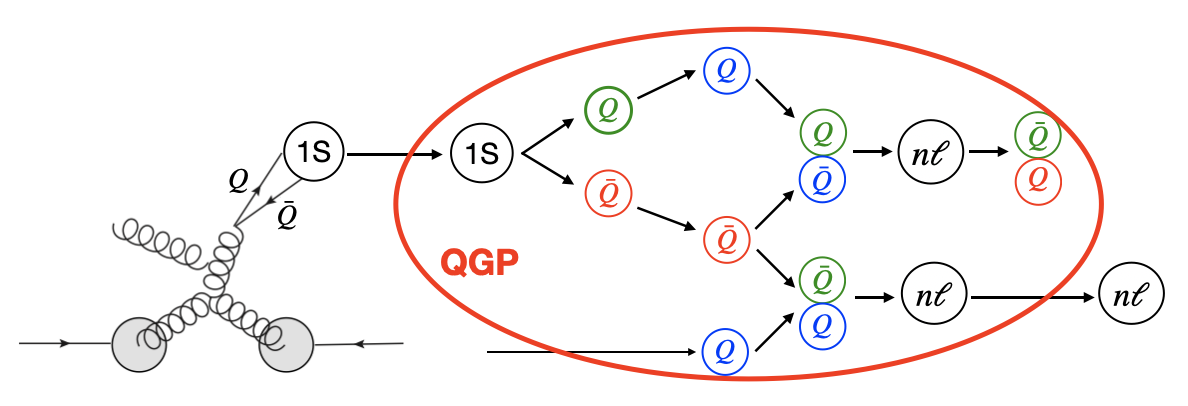}
\caption{Schematic diagram for the dynamics described by the coupled Boltzmann equations.}
\label{fig:coupled_boltzmann}       
\end{figure}

The next example is the set of coupled Boltzmann equations used in the Duke-MIT approach~\cite{Yao:2020xzw}. It combines Boltzmann equations for unbound $Q\bar{Q}$ pairs (effectively for the high temperature region $M\gg T\gtrsim \frac{1}{r}$) and each species of quarkonium states (effectively for the low temperature region $M\gg \frac{1}{r} \gg T,E_b,\Lambda_{\rm QCD}$) with proper collision terms accounting for heavy quark diffusion $\mathcal{C}_{Q\bar{Q}}$, quarkonium dissociation $\mathcal{C}^{-}$, and recombination $\mathcal{C}^{+}$
\begin{align}
&\Big(\frac{\partial}{\partial t} + \dot{{\boldsymbol x}} _Q\cdot \nabla_{{\boldsymbol x}_Q} + \dot{{\boldsymbol x}} _{\bar{Q}}\cdot \nabla_{{\boldsymbol x}_{\bar{Q}}} \Big) f_{Q\bar{Q}}({\boldsymbol x}_Q, {\boldsymbol p}_Q, {\boldsymbol x}_{\bar{Q}}, {\boldsymbol p}_{\bar{Q}}, t) =
\mathcal{C}_{Q\bar{Q}}  -  \mathcal{C}_{Q\bar{Q}}^{+} +  \mathcal{C}_{Q\bar{Q}}^{-} \,,\\
&\Big(\frac{\partial}{\partial t} + \dot{{\boldsymbol x}}\cdot \nabla_{\boldsymbol x}\Big)f_{nl}({\boldsymbol x}, {\boldsymbol p}, t) = \mathcal{C}_{nl}^{+}-\mathcal{C}_{nl}^{-} \,.
\end{align}
In the Boltzmann equation for the unbound $Q\bar{Q}$ pairs, it keeps track of two-particle correlations and does not assume the factorization of the two-particle distribution function, i.e., $f_{Q\bar{Q}}\neq f_Q f_{\bar{Q}}$. Therefore the approach can track both correlated and uncorrelated recombination. In the correlated recombination, the $Q\bar{Q}$ pair comes from the same initial hard collision vertex. In the uncorrelated case, the pair comes from different initial hard collision vertices. Figure~\ref{fig:coupled_boltzmann} schematically shows the dynamics described by the coupled Boltzmann equations.

The Boltzmann equations or rate equations have also been used in the Saclay~\cite{Blaizot:2018oev}, Tsinghua~\cite{Liu:2010ej,Zhao:2022ggw}, TAMU~\cite{Du:2017qkv,Du:2022uvj} and Santiago comover~\cite{Ferreiro:2018wbd} approaches. Each model differs in how they estimate the dissociation rates and recombination contributions. In fact, one does not have to model, but derives them from first principles in certain limits. By using the open quantum system framework and pNRQCD with the hierarchy $M\gg \frac{1}{r} \gg T,E_b,\Lambda_{\rm QCD}$, the Boltzmann and rate equations for quarkonium can be derived in the quantum optical and semiclassical limits~\cite{Yao:2018nmy,Yao:2020eqy}. It was shown that these reaction terms can be factorized into two parts: one part that only knows about quarkonium wavefunction and the other that solely encodes relevant QGP properties. For example, the dissociation rate and the recombination contribution in the rate equation for a quarkonium state $b$ can be written as
\begin{align}
&\frac{{\rm d} n_b({\boldsymbol x},t)}{{\rm d} t} = -\Gamma  n_b({\boldsymbol x},t) + F({\boldsymbol x},t) \,, \\
&\Gamma = \int \frac{{\rm d}^3p_{{\rm rel}}}{(2\pi)^3} | \langle \psi_b | {\boldsymbol r} | \Psi_{{\boldsymbol p}_{\rm rel}} \rangle |^2 [g^{++}_{\rm adj}]^{>}(-\Delta E) \,, \\
&F = \int \frac{{\rm d}^3p_{{\rm cm}}}{(2\pi)^3}  \frac{{\rm d}^3p_{{\rm rel}}}{(2\pi)^3}  f_{Q\bar{Q}} | \langle \psi_b | {\boldsymbol r} | \Psi_{{\boldsymbol p}_{\rm rel}} \rangle |^2 [g^{--}_{\rm adj}]^{>}(\Delta E) \,,
\end{align}
where $\langle \psi_b | {\boldsymbol r} | \Psi_{{\boldsymbol p}_{\rm rel}} \rangle$ is the dipole transition amplitude for the quarkonium state $b$ to an unbound pair with relative momentum ${\boldsymbol p}_{\rm rel}$.
The QGP part is given by chromoelectric correlators at $\Delta E = E_b+\frac{{\boldsymbol p}_{\rm rel}^2}{M}$, which are defined in terms of two chromoelectric fields connected via adjoint Wilson lines
\begin{align}
[g_{\rm adj}^{++}]^>(t) &\equiv \frac{g^2 T_F }{3 N_c}  \big\langle E_i^a(t)W^{ab}(t,0)  E_i^b(0) \big\rangle_T \, , \\
[g_{\rm adj}^{--}]^>(t) &\equiv \frac{g^2 T_F }{3 N_c} \big\langle W^{dc}(-i\beta - \infty, -\infty)
W^{cb}(-\infty,t)  E_i^b(t)
E_i^a(0)W^{ad}(0,-\infty)  \big\rangle_T  \,.
\end{align}
Detailed balance which is used in other models can be shown as a consequence of a field-theory Kubo-Martin-Schwinger (KMS) relation $[g_{\rm adj}^{++}]^>(\omega) = e^{\omega/T} [g_{\rm adj}^{--}]^>(-\omega)$~\cite{Binder:2021otw}. Physically, the chromoelectric correlator can be thought of as a new type of generalized gluon distribution in the medium~\cite{Nijs:2023dbc}. It generalizes all the perturbative QCD diagrams into a gauge-invariant and nonperturbative object: an effective ``gluon'' interacting with the $Q\bar{Q}$ pair in the dipole limit. It is valid to all orders in perturbation theory at leading power in $rT$.

\subsection{Lindblad Equation and pNRQCD}
The last example is the Munich+KSU approach~\cite{Brambilla:2024tqg}. They studied a Lindblad equation in the quantum Brownian motion limit with this hierarchy of energy scales $M\gg\frac{1}{r}\gg T,\Lambda_{\rm QCD}$; $T \gg E_b$
\begin{align}
\frac{{\rm d} \rho_{Q\bar{Q}}(t)}{{\rm d} t} = -i\big[ H + \gamma_{\rm adj} \Delta h,\, \rho_{Q\bar{Q}}(t) \big] +  \kappa_{\rm adj} \Big( L_{\alpha i} \rho_{Q\bar{Q}}(t) L^\dagger_{\alpha i} - \frac{1}{2}\big\{  L^\dagger_{\alpha i}L_{\alpha i},\, \rho_{Q\bar{Q}}(t)\big\} \Big) \,.
\end{align}
The Lindblad equation depends on two new generalized transport coefficients $\kappa_{\rm adj}$ and $\gamma_{\rm adj}$, which can be defined in terms of the chromoelectric correlator
\begin{align}
\kappa_{\rm adj} &\equiv [g_{\rm adj}^{++}]^>(\omega=0) = \frac{g^2T_F}{3N_c} \int{\rm d}t \, \big\langle E_i^a(t) W^{ab}(t,0) E_i^b(0) \big\rangle_T \\
\gamma_{\rm adj} &\equiv \frac{ g^2 T_F}{3 N_c} {\rm Im} \!\int {\rm d} t \, \big\langle \mathcal{T} E^a_i(t) W^{ab}(t,0) E^b_i(0) \big\rangle_T  \,.
\end{align}
In their most recent work~\cite{Brambilla:2024tqg}, they used the three-loop QCD potential, which can describe the bottomonium spectrum in vacuum more accurately than using the Coulomb potential.

\section{Chromoelectric Correlator}
I hope I have convinced you that in many transport studies for quarkonium, it is the chromoelectric correlator that encodes the relevant properties of the QGP. The heavy quark diffusion coefficient is also defined by a chromoelectric correlator with fundamental Wilson lines
\begin{align}
\label{eqn:kappa_fund}
\kappa_{\rm fund} = \frac{g^2}{3N_c} {\rm Re}\int {\rm d} t\, 
\big\langle {\rm Tr}_{\rm c}[ U(-\infty,t) E_i(t) U(t,0) E_i(0) U(0, -\infty)] \big\rangle_{T,Q} \,.
\end{align}
But the correlator for heavy quark diffusion in Eq.~\eqref{eqn:kappa_fund} and that for quarkonium, i.e., $[g_{\rm adj}^{++}]^>$ are two different physical objects. They have different operator orderings. Their difference has also been evaluated in perturbation theory~\cite{Eller:2019spw,Binder:2021otw}. The difference is nonzero at next-to-leading order (NLO) in Feynmann gauge. However, in temporal (axial) gauge, the two correlators would look identical. This is the temporal (axial) gauge puzzle, first noted by Eller, Ghiglieri and Moore~\cite{Eller:2019spw}. However, they did not solve the puzzle. It was Scheihing-Hitschfeld who solved it~\cite{Scheihing-Hitschfeld:2022xqx}. The crucial observation is that one cannot smoothly define the temporal (axial) gauge for the heavy quark diffusion correlator. The calculation in the temporal (axial) gauge gives the quarkonium correlator result. The quarkonium correlator was further calculated in a strongly coupled supersymmetric Yang-Mills plasma~\cite{Nijs:2023dks}. The comparison with the perturbative QCD result can be found in Ref.~\cite{Nijs:2023dbc}. The results are normalized to agree at large positive frequency. At negative frequency, the correlator decreases as the coupling increases. In the strong coupling limit, it vanishes. Physically, it means both $\kappa_{\rm adj}$ in the Lindblad equation and $[g^{\pm\pm}_{\rm adj}]^>$ in the Boltzmann equation vanish: These evolution equations become trivial. In other words, Markovian dynamics of quarkonium becomes trivial in the strong coupling limit. This motivates us to think about non-Markovian dynamics of quarkonium. Progress has been reported in this conference by Scheihing-Hitschfeld~\cite{bruno}. 

\section{Conclusions and Future Prospects}
OK, now let me recap. In this talk, I showed you how low-$p_T$ data of quarkonia can help us to understand the chromoelectric correlator of the QGP. It is not an easy task. We need theory developments, phenomenological studies and computational advancements. 

It is important to discuss where this field is going. Here is a list of some directions that I think are worth mentioning.

The first question is the microscopic structure of the QGP probed by various quarkonium states. For example, $\Upsilon(nS)$ has three (well-defined) bound states, thus providing three scales to probe the QGP. Is the QGP a weakly coupled gas of quarks and gluons or a strongly coupled fluid at the scale probed by $\Upsilon(1S)$? Similarly for $\Upsilon(2S)$ and $\Upsilon(3S)$? We want to pin down the chromoelectric correlator from experimental data. To do this, we need both LHC and RHIC data. I want to emphasize that the RHIC data is important to determine the frequency dependence of the chromoelectric correlator, since the dynamics is more sensitive to it at low temperature. This is a great opportunity for the sPHENIX program. 

The second question is the nonperturbative determination of the chromoelectric correlator $[g^{\pm\pm}(\omega)]^>$ from QCD. After its extraction from experimental data, we need to compare with perturbative QCD and nonperturbative results in order to decipher the microscopic structure of the QGP. There have been studies attempting to do the nonperturbative calculation via the lattice QCD method. Two crucial aspects of the calculation are the non-oddness of the associated spectral function~\cite{Scheihing-Hitschfeld:2023tuz} and the renormalization of the correlator~\cite{Scheihing-Hitschfeld:2023tuz,Brambilla:2023vwm,Leino:2024pen}.

The third question is about the spin polarization of quarkonium. We have seen new experimental data for this and two studies have discussed the relevant phenomenology~\cite{Cheung:2022nnq,Zhao:2023plc}. One can repeat what I have just presented in this talk, apply the open quantum system framework and pNRQCD and show that quarkonium polarization measurements probe the chromomagnetic correlator $\big\langle B_i^a(t) W^{ab}(t,0) B_i^b(0) \big\rangle_T$~\cite{Yang:2024ejk}. It is another unique property of the QGP that can be probed by quarkonium.

Finally, I want to mention novel techniques to solve the Lindblad equation. In this talk, all the Lindblad equation studies I mentioned solved for just one $Q\bar{Q}$ pair. We know that for charmonium, multi-charm uncorrelated recombination is an important contribution to the $J/\psi$ production. Solving the Lindblad equation for multiple $Q\bar{Q}$ pairs is computationally very expensive. So we may seek new methods such as machine learning~\cite{Lin:2024eiz} or quantum computing~\cite{DeJong:2020riy,deJong:2021wsd}. These ideas have been tested in simple models. Applications to quarkonium dynamics will come. Stay tuned.\\

This work is supported by the U.S. Department of Energy, Office of Science, Office of Nuclear Physics, InQubator for Quantum Simulation (IQuS) (https://iqus.uw.edu) under Award Number DOE (NP) Award DE-SC0020970 via the program on Quantum Horizons: QIS Research and Innovation for Nuclear Science.

%
%

\end{document}